\documentclass[aps,12pt]{revtex4}

\usepackage{epsfig}
\usepackage{graphicx}
\usepackage{amsmath,amssymb,color}
\usepackage[english]{babel}
\usepackage{latexsym}
\usepackage{amsfonts}
\usepackage{natbib}
\usepackage{ulem}
\usepackage{float}

\parskip=\medskipamount

\newcommand{\eq}[1]{(\ref{#1})}
\newcommand{\fig}[1]{Fig. \ref{#1}}

\newcommand{\be}{\begin{equation}}
\newcommand{\ee}{\end{equation}}

\newcommand\disp{\displaystyle}

\newcommand{\la}{\left<}
\newcommand{\ra}{\right>}

\begin{document}

\title{Eigenvalue detachment, BBP transition and constrained Brownian motion}

\author{Alexander Gorsky$^{1}$, Sergei Nechaev$^2$, and Alexander Valov$^3$}

\affiliation{$^1$Institute for Information Transmission Problems RAS, 127051 Moscow, Russia \\ 
$^2$LPTMS, CNRS -- Universit\'e Paris Saclay, 91405 Orsay Cedex, France\\ 
$^3$N.N. Semenov Federal Research Center for Chemical Physics RAS, 119991 Moscow, Russia}

\date{today}

\begin{abstract}
We discuss the eigenvalue detachment transition in terms of scaling of fluctuations in ensembles of paths located near convex boundaries of various physical nature. We consider numerically the BBP-like (Baik-Ben Arous-P\'ech\'e) transition from the Gaussian to the Tracy-Widom scaling of fluctuations in several statistical systems for both canonical and microcanonical ensembles and identify the corresponding control parameter in each case. In particular, for fixed path length (microcanonical) ensemble of paths located in the vicinity of a partially permeable semicircle, the  transition occurs at the critical value of a permeability. The Tracy-Widom regime and the BBP-like transition for fluctuations are discussed in terms of the Jakiw-Teitelbom (JT) gravity with a radial cutoff which, in turn, has an interpretation as a ensemble of fixed length world-line geometrically constrained trajectories of a charged particle in an effective transversal magnetic field.

\end{abstract}

\maketitle

\section{Introduction: General frameworks of unconventional transitions}

Scaling laws of fluctuations provide the convenient instrument of probing statistical behavior of the system. Two conventional laws, ``Gaussian'' and ``Tracy-Widom'' manifest themselves in a huge variety of statistical systems. The common Gaussian law is typically related with the central limit theorem for a large number of independent random variables, while the Tracy-Widom (TW) law \cite{tracy2009distributions} emerges in extreme statistics of a large number of correlated random variables. The behavior of some observables in scaling regimes is governed by universal solutions of differential equations. For example, the solution to the Painleve II yields the scaling function of the TW regime. It has been explicitly shown in \cite{baik2005phase} that the eigenvalue detachment can be microscopically identified with the transition for the largest eigenvalue, $\lambda_{max}$, from the TW distribution (when $\lambda_{max}$ correlates with the bulk) to the Gaussian one (when the correlations of $\lambda_{max}$ with the bulk are absent). The new ``intermediate'' distribution emerges exactly at the transition point.

The detachment of the largest eigenvalue from the bulk of the spectrum is the phenomena which can be seen in various physical situations. Historically such a detachment has been first discussed in detail in the context of spin glasses \cite{kosterlitz1976spherical}. In this case the control parameter was identified with the eigenvalue of the matrix which centers the Gaussian ensemble with the shifted mean. More recently, this phenomenon was considered in the probability theory for the behavior of eigenvalues of some covariance matrix \cite{baik2005phase}. The control parameter is the deviation of a single eigenvalue of the covariance matrix from the unity. Later more general situation with several control parameters was discussed as well. Exactly at the transition point the new universality class has been identified which distribution which now is known as the ``Baik-Ben Arous-P\'ech\'e'' (BBP) distribution. Similar to the Tracy-Widom (TW) distribution which is given by the solution to the Painleve II equation with the specific {\it asymptotics}, the BBP distribution is described by the pair of differential equations which involve the solution to Painleve II equation with the peculiar {\it monodromy} \cite{baik2006painleve}.

The shifted mean Gaussian ensemble has been generalized to the chiral case in \cite{bassler2009eigenvalue} and the position of the separated eigenvalue as a function of the order parameter has been derived analytically. Later on more eigenvalue detachment phase transitions were found in such physical problems as: last passage percolation \cite{lpp}; Asymmetric Simple Exclusion Process (ASEP) with particular spiked initial conditions \cite{borodin2014free,aggarwal2019phase}; $q$-version of Totally Asymmetric Simple Exclusion Process ($q$-TASEP) with slow particles \cite{barraquand2015phase}; percolation in 2D \cite{saber2022universal}; spin glass-paramagnetic transition in the mean field approximation \cite{glass}. In all these cases the nature of the phenomena is one and the same: first one zooms the spectral edge where the TW distribution emerges, and then the perturbation is introduced. 

In our study we focus our attention not at the distribution of spectral fluctuations, rather at the distribution of paths' fluctuations in space-time. The fermionic nature of eigenvalues in the matrix model can be mapped onto the vicious walkers problem. So, the detachment of one eigenvalue from the bulk of the spectrum gets mapped onto the emergence of the outliers in the Brownian ensemble. The dynamics of the Brownian motion of large number of walkers with a few outliers has been also discussed in the mathematical literature in \cite{adler2009dyson} where the transition from the Gaussian to TW fluctuations has been derived analytically. Instead of looking at detached eigenvalues, the same pattern can be modelled in the ensemble of individual random walkers nearby the extended defect. In the polymer language, the BBP transition occurs when a (1+1)D random walk is located exactly at the threshold of the formation of a bound state with some type of defect, or attractive boundary. As shown in examples below, the interaction of a random path with a defect can be introduced either explicitly of effectively.

The example of an explicit interaction is provided in the work \cite{krajenbrink2021tilted}, where the BBP-like transition emerges when a part of a polymer trajectory gets localized on an extended defect at some critical value of coupling between the polymer and the defect. Mapping of the polymer problem onto the matrix model in the context of the BBP transition has been discussed in \cite{krajenbrink2021tilted}. Conceptually the model under consideration is as following. Take a random matrix playing a role of a transfer-matrix for some lattice model and suppose that the spectrum of this matrix shares the Wigner semicircle law (that is a rather generic supposition). It is known that at the edge of the semicircle the largest eigenvalue has the Tracy-Widom distribution. Let one deform the transfer-matrix by a perturbation involving some coupling constant, $u$. Above a critical coupling, $u_c$, one eigenvalue detaches from the continuum, which is the ``spectral manifestation'' of the BBP transition.  

The effective interaction of a path with an extended defect has been studied in \cite{nechaev2019anomalous,vladimirov2020brownian,gorsky2018statistical,valov2021equilibrium} in the context of the Ferrari-Spohn problem \cite{ferrari2005constrained}. Being rephrased in polymer terms, the problem is as follows: the part of a polymer trajectory located in the vicinity of a convex void boundary experiences the transition between different fluctuation regimes as a function of a boundary curvature. The corresponding behavior has been interpreted in \cite{meerson2019geometrical,smith2019geometrical} as a ``shadowing'' a path by a convex impermeable boundary on which the path is leaning. In such a setup the key point is the consideration of the {\it microcanonical} ensemble of fluctuating paths of a fixed length, $N$. The condition $N=cR$ is imposed on paths with ends fixed at opposite extremities of a diameter of a convex void (a semicircle of radius $R$). Varying $c$, which is the control parameter, one sees the transition from TW scaling to the Gaussian one, which we interpret as a BBP transition. The transition here is induced by fixing the path's length which forces the trajectories to nestle against the disc boundary and $c$ is the control parameter which governs the strength of the ``pinching force'' and plays a role of an effective curvature of a disc's boundary for a path of a given length. 

In the present work we discuss numerically manifestations of the BBP transition in various scenarios for a single random path nearby the convex boundary. Specifically, we consider several formulations:
\begin{itemize}
\item [(i)] Microcanonical ensemble of random paths of fixed length, $L=cR$, above an {\it impermeable} disc of radius $R$ where $c$ is the control parameter for BBP transition
\item[(ii)] Microcanonical ensemble of  random paths of fixed length $L=cR$, above a {\it partially impermeable} disc of radius $R$, in which the fraction of chain monomers inside the disc is controlled by the parameter $\eta$. We see that $\eta$ is the control parameter for the BBP transition;
\item[(iii)] Canonical ensemble of random paths whose length, $L$, is controlled by the chemical potential $s$ above the {\it fish-like defect}. The angle at the cusp is the control parameter.
\end{itemize}

The interplay between the BBP transition in terms of spectral and spatial fluctuations can be naturally understood in the holographic framework. The radial coordinate in the hyperbolic plane has the meaning of the energy scale in the boundary theory. Hence indeed the BBP transition in the scaling regime of fluctuations for eigenvalues in the boundary theory fits with the radial fluctuations in its holographic dual. Since the BBP transition concerns the maximal eigenvalue, the introduction of some radial cutoff is expected. We shall comment on the TW regime and the BBP transition in Jakiw-Teitelbom (JT) gravity holographically described via the quantum mechanics with large number degrees of freedom. The key point is the identification of the partition function of JT gravity as the function of the ensemble of the fixed length paths of a charged particle in the hyperbolic plane in the transversal magnetic field \cite{kitaev2019statistical,yang2019quantum}. To get the TW regime we should introduce the radial cutoff $R_{cut}$ which corresponds the energy cutoff in the boundary theory and then tune the temperature is such way that the cutoff radius approaches the Larmour radius $R_{Lar}$ of a charged particle in the transversal magnetic field. In this case we arrive at the framework which is similar to one discussed for the bunch of trajectories nearby the hard circle (which mimics the cutoff radius). The quotient $R_{cut}/R_{Lar}$ is the control parameter in this case.

The paper is structured as follows. In Section II we recall the general framework of the BBP phase transition. In Section III we consider different formulations of the model and observe numerically  the transition from KPZ to Gaussian regime for fluctuations in each case. In Section IV we formulate the Tracy-Widon regime and the BBP-like transition in the JT gravity with the radial cutoff under particular limitation of parameters. In Section IV we summarize obtained results and speculate about possible further developments. In Appendix A we consider the trajectory in the background of an attractive defect. Such a scenario has been discussed for the flat defect in \cite{krajenbrink2021tilted} while we have presented another example of a similar kind for the convex defect. Considered setting describes the induced false vacuum decay in the (1+1)-dimensional space-time. In Appendix B we use the nonlinear UMAP method of the dimensional data reduction to relate the BBP transition with the changes of the data spot in the abstract two-dimensional plane. In Appendix C we present for completeness some known formulae concerning the BBP transition.

\section{Tracy-Widom scaling law for fluctuations and Baik-Ben Arous-Pech\'e transition}

Here we briefly recall the origin of the Tracy-Widom (TW) distribution and the formulation of the Baik-Ben Arous-P\'ech\'e (BBP) transition for fluctuational statistics. Historically the Tracy-Widom law has been identified for the Airy kernel in some random matrix models
\be
A(x,y)= \int _{0}^{\infty}{\rm Ai}(z+x){\rm Ai}(z+y) dz = \frac{{\rm Ai}'(x){\rm Ai}(y) - {\rm Ai}'(y){\rm Ai}(x)}{x-y}
\ee
The TW distribution, $F_0$, can be expressed in terms of the Fredholm determinant
\be
F_0(s)= \det\Big(1- A_s(x,y)\Big)
\ee
where $A_s(x,y)$ is the corresponding kernel operator. The TW distribution can be expressed in terms of solutions of the Painleve II equation as follows
\be
F_0(s)=\exp\left(- \int_{x=s}^{\infty} (s-x)^2u^2(x)dx\right)
\ee
where function $u(x)$ obeys the Painleve II equation
\be
u''(x)=2u^3(x) +x u(x)
\label{eq:pain2}
\ee
subject to the specific boundary condition defined by the asymptotics
\be
u(x)\propto -{\rm Ai}(x), \quad x\rightarrow{+\infty}
\ee
The function $F_0(s)$ describes the fluctuations of the largest eigenvalue $\lambda_{max}$ in ensembles of interacting particles. For instance for the Laguerre unitary ensemble one has
\be
P\Big(\lambda_{max} - c N^{2/3}\leq s\Big) \rightarrow  F_0(s)
\ee
where $c$ is parameter of the model.

It is known that the Tracy-Widom scaling is also provided by the solutions to the one-dimensional KPZ equation for the height function $h(x,t)$ developing in time $t$:
\be
\frac{\partial h(x,t)}{\partial t}= \frac{1}{2} \left(\frac{\partial h(x,t)}{\partial x}\right)^2 + \frac{1}{2} \frac{\partial^2 h(x,t)}{\partial x^2} + W(x,t)
\label{eq:kpz}
\ee
where $W(x,t)$ is the white noise in (1+1)D space-time. Upon the Cole-Hopf transform $Z=e^{h(x,t)}$, Eq.\eq{eq:kpz} it can be brought into the form
\be
\frac{\partial Z(x,t)}{\partial t}=  \frac{\partial^2 Z(x,t)}{\partial x^2} + W(x,t) Z(x,t)
\ee
where the function $Z(x,t)$ can be interpreted as the partition function of the polymer in the plane in an external random potential. At large $t$ the height function behaves as
\be
h(0,t) \propto -\frac{t}{24} + \left(\frac{t}{2}\right)^{1/3}\eta
\ee
where $\eta$ for the wedge initial condition is the random Tracy-Widom distributed variable coinciding with the distribution of the largest eigenvalue of the Gaussian Unitary Matrix Ensemble (GUE) at large matrix sizes, $N$. The distribution is non-universal for KPZ and depends on the initial condition in a KPZ equation. 

In what follows we shall be interested in the interpretation of $Z(x,t)$ as of the partition function of $N$-step directed random walks (polymers) in a random media with the initial condition $Z(x,0)=\delta(x)$ and ending point located at $(x,t)$. The large-$N$ behavior of the free energy of ensemble of such polymers can be derived from KPZ equation and reads \cite{dotsenko,doussal-kpz}:
\be
F = T \ln Z(0,N) \propto -N E_0 + a N^{\frac{1}{3}}
\ee
where $E_0$ and $a$ are some constants. Upon the Laplace transform of $Z(0,N)$, the Lifshitz tail for the spectral density can be recovered \cite{gorsky2021lifshitz}. In our study the transition between two different scaling behaviors of fluctuations: the Gaussian-like and the Tracy-Widom-like will be discussed for a single random walker near a convex surface.

The BBP distribution (or ``spiked TW law'') emerging exactly in the transition regime can be also expressed in terms of the solution to the Painleve II equation. The scaling function for the case when the single eigenvalue gets detached from the bulk has the form 
\be 
F_1(s,\omega)= F_0(s) f(s,\omega)
\label{eq:f1}
\ee
where the new function $f(s,\omega)$ obeys the system of second order differential equations \cite{baik2006painleve} for a pair of unknown functions, $f$ and $\omega$:
\be
\begin{cases}
\disp -f_{ss} +\left(\frac{u_{s}}{u} -\omega\right)f_{s} + u^2 f =0 \medskip \\
\disp -f_{\omega \omega} + \left(\frac{u}{\omega u + u_{s}} +\omega^2 -s\right)f_{\omega} + \left(u^4 +
su^2 - u_{s}^2 - \frac{u^3}{\omega u + u_{s}}\right)f=0
\end{cases}
\label{eq:pair}
\ee
where $u(s)$ is solution to the Painleve II equation \eq{eq:pain2}. The imposed boundary conditions are as follows
\be
f(s,0) = E(s), \qquad f_{\omega} (s,0)= (u^2 +u_s)E(s)
\ee
and
\be
E(s)= \exp\left(\int_{s}^{\infty} u(s) ds\right) 
\ee
The solution of \eq{eq:pair} can also be expressed in terms of the $2\times 2$ monodromy matrix for the Riemann-Hilbert problem of the Painleve II equation when some additional constraints are imposed on the monodromy data \cite{baik2006painleve}. The monodromy problem subject to this constraint has an unique solution. The explicit expressions for more general scaling function $F_k(s,\omega_1 \dots \omega_n)$ when $n$ eigenvalues get detached from a bulk of the spectrum is known as well and can be expressed by a straightforward generalization of \eq{eq:pair}. The determinantal representations for these scaling functions are also available \cite{baik2006painleve}.

\section{Random walks above the convex void: various boundary conditions and various ensembles}

\subsection{Microcanonical ensemble of random paths evading fully impermeable semicircle}

Despite the tremendous progress in understanding the mathematical background of the Tracy-Widom distribution and its relation to third-order phase transitions \cite{majumdar2014top}, still there is an essential lack in constructing clear and simple statistical models of the mean-field nature which, on one hand, share the KPZ-type scaling for fluctuations and, on the other and, clearly demonstrate the emergence of the third order phase transition. In connection with that a very promising is the systems proposed by H. Spohn and P. Ferrari in \cite{ferrari2005constrained} where the statistics of one-dimensional directed random walks evading the semicircle has been discussed. It is known that the fluctuations of the top line in a bunch of $n$ one-dimensional directed ``vicious walks'' glued at their extremities (fermionic world lines in 1D) are governed by the TW distribution. Proceeding as in \cite{ferrari2005constrained}, one can define the averaged position of the top line and look at its fluctuations. In such a description, all vicious ``bulk'' walks lying below the top line, play a role of a ``mean field'' and are pushing the top line to some new equilibrium position. Fluctuations around this position differ from fluctuations of a free random walk in absence of the bulk. Replacing the effect of the bulk by the semicircle of radius $R$, one arrives at the Spohn-Ferrari model where the 1D directed Brownian bridge stays above the semicircle, and its interior is inaccessible for the path. 

The bunch of works \cite{meerson2019geometrical,smith2019geometrical} extending the Ferrari-Spohn model \cite{ferrari2005constrained} have made a significant contribution to the development of such models. These papers provided transparent ``geometric optic'' approach for statistics of large deviations of Brownian trajectories pushed by external geometric constraints to an atypical region of the phase space. In \cite{meerson2019geometrical} it has been shown that the transition from the KPZ-like to the Gaussian behavior occurs when the path fluctuating above the impermeable disc is long enough with respect to the disc boundary, to have a freedom to escape from the ``shadow'' produced by the disc. 

\begin{figure}[ht]
\centering
\includegraphics[width=.95\textwidth]{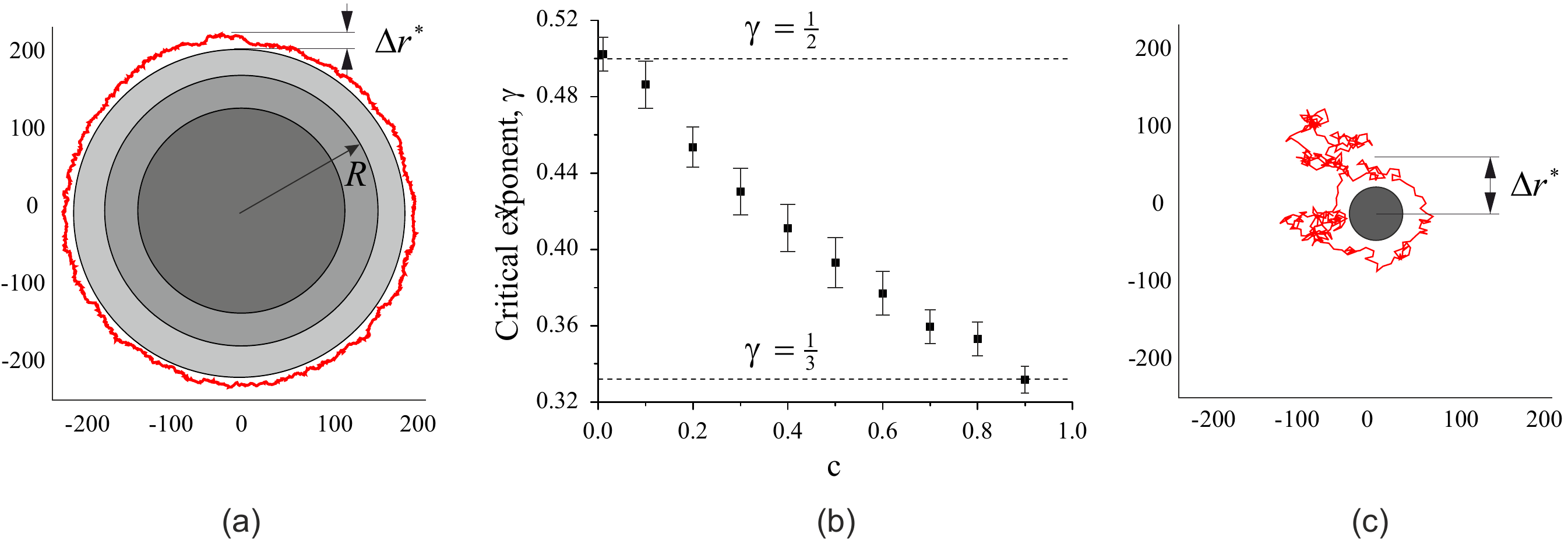}
\caption{(a) Polymer loop of length $N=7R_g$ leaning on an impenetrable disc of changing radius $R=cR_g$; (b) Dependence of the critical exponent $\gamma$ on $c$, where $\gamma(c)$ is defined  as $\Delta r^*(N) \propto N^{\gamma(c)}$; (c) Typical snapshot of the system for $c\approx 0.05$.}
\label{fig:001}
\end{figure}

In our study we consider the microcanonical ensemble of the random directed paths of a length $N=cR$ ($c={\rm const}$), and fixed boundary conditions, which stay in the vicinity of an impermeable disc of radius $R$ such that $\pi R < N \ll R^2$. Varying the parameter $c$, we observe transition for the scaling of fluctuations from the Gaussian behavior to the KPZ ones -- see \fig{fig:001}. It it important to emphasize that the transition takes place {\it along the whole trajectory}. Thus, the immediate question emerges concerning the interpretation of $N$ in the KPZ-type scaling. Indeed the KPZ ``1/3'' scaling law corresponds to the large-time asymptotics, hence interpretation of $N$ as of a ``time'' variable is impossible. The correct answer seems as follows: due to the imposed relation between $N$ and $R$ ($N=cR$), the variable $N$ is changing by changing $R$ and the evolution of our system can be considered in the radial coordinate, which plays a role of an ``renormalization group''-like coordinate in the holographic approach. The scaling behavior of the span $\Delta \sim R^{\gamma}$ of random paths above the impermeable disc as a function of $R$ has been checked numerically for different values of the parameter $c$, and indeed at large $R$ the KPZ-like scaling is well reproduced -- see the corresponding plot in \fig{fig:001}. 

\subsection{Microcanonical ensemble of stretched paths evading partially impermeable semicircle}

\begin{figure}[ht]
\centering
\includegraphics[width=.95\textwidth]{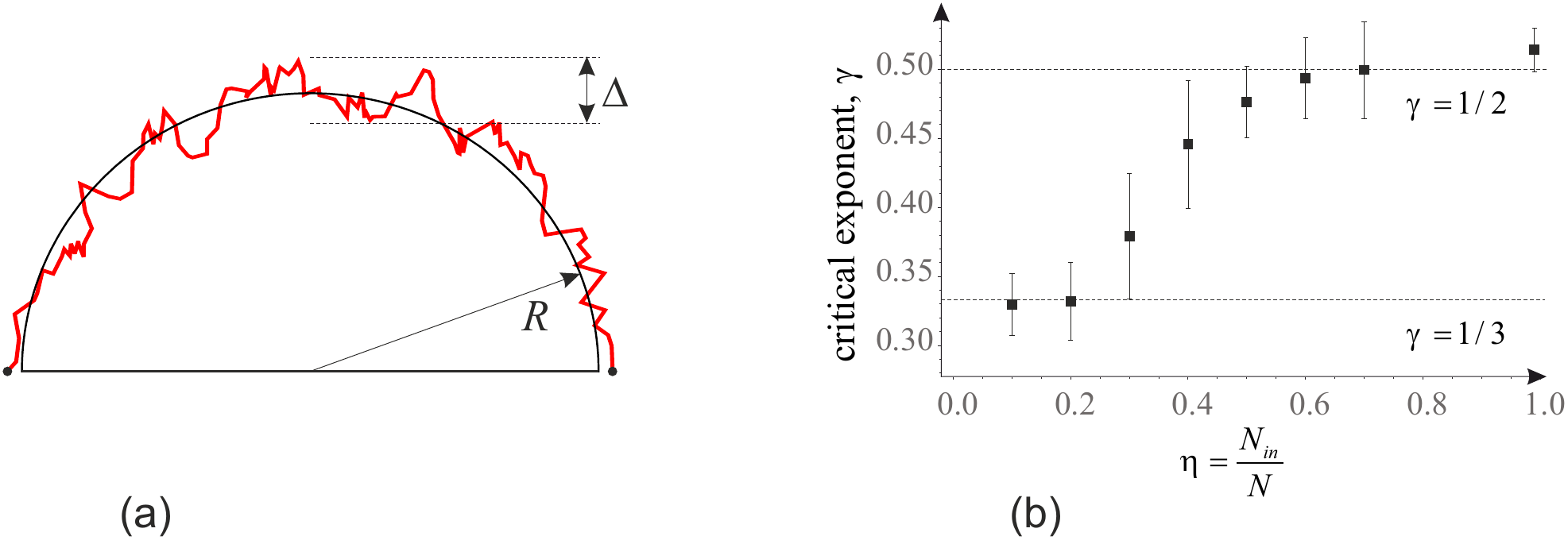}
\caption{(a) Sample of the path of length $N=4 R$, that envelopes a partially impermeable semicircle of radius $R=30$ for $\eta=0.7$; (b) Phase transition curve $\gamma(\eta)$, where $\gamma$ is the scaling exponent in the dependence $\Delta(N)\sim N^{\gamma}$ for strong path stretching as in (a).}
\label{fig:01}
\end{figure}

Let us modify a model discussed above and assume that the disc is semi-permeable for paths as it is shown in \fig{fig:01}. The probability for a path to stay inside the disc we denote as $\eta$. Varying $\eta$ from $\eta=0$ (all paths stay outside of the disc) to $\eta=1$ (all paths stay inside the disc), we probe fluctuations of fixed length paths, which depend on direct contact interactions of paths with the disc boundary. The value $\eta$ plays a role of an ``order parameter'' for a system under consideration. 

Considered setup mimics to some extent the transition from ensembles of fermionic world lines to  ensembles of interacting ``anyonic'' Brownian walkers. Namely, as it has been said above, the completely impermeable disc inaccessible for an exterior trajectory effectively describes a system of vicious Brownian walkers aka fermionic worldlines. Meanwhile, if a disc is partially impermeable, the exterior path could penetrate inside a bulk of a disc. This situation effectively describes a system of ``semi-vicious'' Brownian walkers possessing the anyonic statistics. At $\eta=1$ the Brownian walker does not feel a disc at all and we have reached the state of non-interacting walkers.

It seems instructive to recall some qualitative arguments involving the naive dimensional analysis. An unconstrained $N$-step random walk, at $N \gg R^2$ fluctuates freely and almost does not feel the constraint (the boundary of the disc). Thus, the only possible scaling for the typical span of the unconstrained random walk is $\Delta \sim N^{1/2}$. In the opposite regime, $\pi R< N \ll R^2$, the chain statistics is essentially perturbed by the disc. In the limit of strong stretching, the suitable width of a ``tube'' within which the path is localized, is given by the scaling relation
\be
\Delta \sim N^{2/3} R^{-1/3}
\label{eq:Delta}
\ee
which under the condition of extreme stretching ($R = N/c$), is reduced to $\Delta\sim N^{1/3}$, while under the condition of unconstrained fluctuations ($R\sim \sqrt{N}$) provides the relation $\Delta\sim N^{1/2}$. Equation \eq{eq:Delta} can be obtained via the ``optimal fluctuation'' method proposed in \cite{grosberg}, in which the free energy, $F(\Delta)$, of an ensemble of stretched paths consists of two terms: (i) stretching, 
$$
F_{str}(\Delta) \sim \frac{(R+\Delta)^2}{N}
$$
and (ii) confinement, 
$$
F_{conf}(\Delta) \sim \frac{N}{\Delta^2}
$$ 
Expanding $F_{str}(\Delta)$ for $\frac{\Delta}{R}\ll 1$ and minimizing $F(\Delta)$, where  
$$
F(\Delta) = F_{str}(\Delta) + F_{conf}(\Delta) = F(\Delta)\sim \frac{R\Delta}{N} + \frac{N}{\Delta^2}
$$ 
with respect to $\Delta$, we arrive at an ``optimal'' $\Delta$ given by \eq{eq:Delta}.

Returning to the model under discussion, let us remind that the parameter $c$ in the relation $N=cR$ is kept fixed, and the disc is semi-permeable for paths. The fraction of monomers inside the disc, $N_{in}$, is characterized by the parameter $\eta=N_{in}/N$. Varying $\eta$ we can pass from the situation in which all trajectories stay outside of the disc ($\eta=0$) to the situation in which the trajectories do not feel the disc boundary at all ($\eta=1$). In terms of the particle trajectories in the Euclidean (1+1)D space-time, the parameter $\eta$ can be considered as a quotient of masses of a particle inside and outside the disc boundary. The corresponding partition function of the microcanonical ensemble of paths can be written as
\be
Z(cR,\eta)=\sum_{{\rm N-step~paths}} \delta(N-cR) \delta\left(\eta - \frac{N_{in}({\rm path})}{N}\right)
\label{eq:01}
\ee
where $N_{in}({\rm path})$ is the number of steps of a specific paths inside a circle, $N$ is the total number of steps.

The setting of our numeric simulations is as follows. Each (1+1)D path of length $N=cR$, wound around a ``semi-permeable'' disc of radius $R$, is represented by a sequence of points $\{(x_i,y_i)\}_{i=[1,N]}$, whose extremities are fixed near opposite sides of an impermeable semicircle, $(-R-\epsilon,0)$ and $(R+\epsilon,0)$, where $\epsilon/R\ll 1$. The distances between neighboring points of the path are Gaussian-distributed random variables with the unit mean and variance. The span of fluctuations of the ensemble of trajectories in the radial direction, $\Delta$, depends on the ``permeability'' $\eta=N_{in}/N$. The limiting case $\eta\to 0$ (i.e. $N_{in}=0$) reproduces the Ferrari-Spohn model, in which trajectories completely evade the semicircle, while the opposite limit, $\eta\to 1$, corresponds to paths entirely staying inside the semicircle except terminal points which by definition are fixed outside the disc. Thus we expect the following scaling dependence, $\Delta(N)\sim N^{\gamma(\eta)}$, which is as follows:
\be
\Delta(N)\sim \begin{cases} N^{1/3} & \mbox{for $0\le \eta < \eta_{cr}$} \medskip \\ N^{1/2} & \mbox{for $\eta_{cr}<\eta < 1$} \end{cases}
\label{eq:02}
\ee
where $\eta_{cr}$ is the transition point which separates two different scaling laws in \eq{eq:02}: $\gamma(\eta<\eta_{cr})=1/3$ and $\gamma(\eta>\eta_{cr})=1/2$.

\subsection{Canonical ensemble of  paths above a ``fish''}

\begin{figure}[ht]
\centering
\includegraphics[width=.95\textwidth]{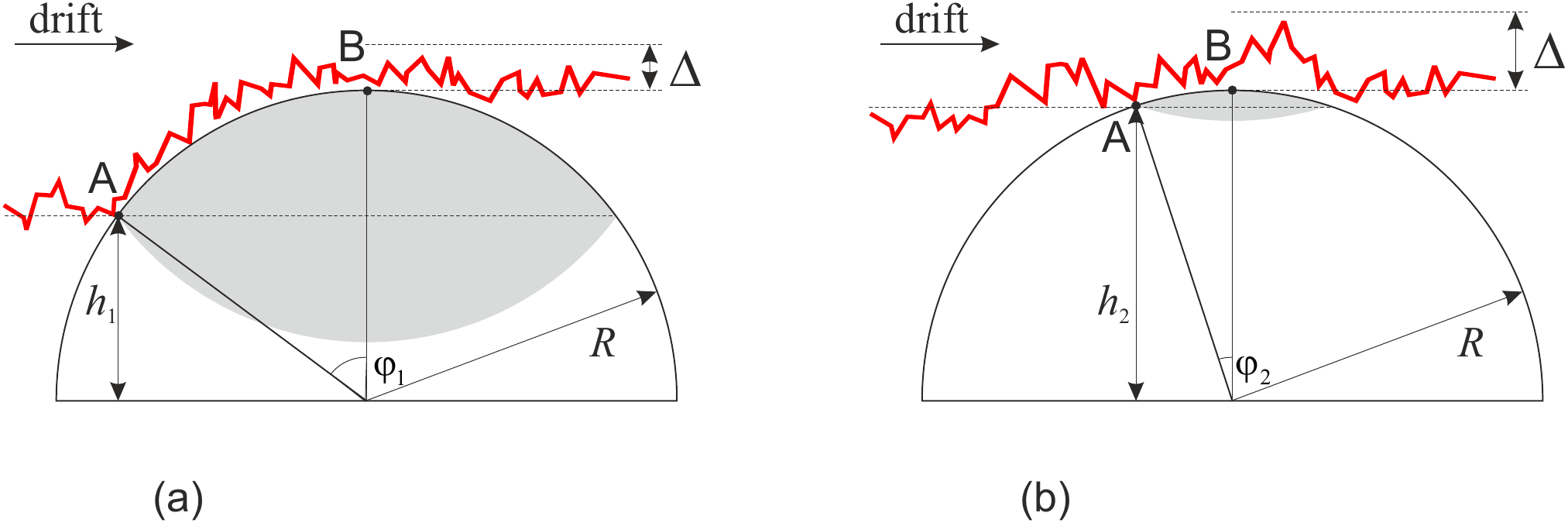}
\caption{Biased trajectories of a tracer particle hitting a circular boundary at some height $h$, the ``fish'' is shown in gray color: (a) long running time form $A$ to $B$ ensures KPZ-like scaling for span $\Delta$ of fluctuations above the boundary, (b) short running time from $A$ to $B$ is not sufficient to change the fluctuational statistics of $\Delta$ from Gaussian.}
\label{fig:02}
\end{figure}

Here we study fluctuations of a tracer particle in a two-dimensional plane near the impenetrable circular boundary as it is schematically shown in \fig{fig:02}. We consider a ``canonical'' setup (length of trajectories is fixed by a chemical potential). Such a setting is typical for the so-called \textit{constant force} active micro-rheology \cite{poon}. The tracer experiences random thermal forces due to interactions with the medium, which acts as a heat bath and is under the action of a constant force pointing along the $x$-axis. Hence, in our settings, the tracer performs an unbiased diffusion in a $y$-direction, perpendicular to an applied force, and a biased diffusion along the field (i.e. along the $x$-axis). Evidently, the longitudinal and normal components of the tracer's motion at any instantaneous position in the vicinity of the boundary are effectively coupled due to interactions with a part of a curved boundary called a ``fish''. Using heuristic arguments, we analyse an impact of these interactions on the fluctuational behavior of the tracer particle.

Specifically, the question concerns the determination of the scaling law of a typical span, $\Delta$, in the dependence $\Delta(t)\sim t^{\gamma}$, as a function of time, $t$, spent by tracer particle between its first touch of the boundary at the point $A$ and its arrival at the point $B$ above the tip of the circular boundary (see \fig{fig:02} for details). Our analysis reveals that, depending on the position of a starting point, which can be characterized by the angle $\phi$ (or equivalently by the height $h$), the scaling exponent, $\gamma$, can experience the transition form $\gamma=1/3$ (the KPZ-like statistics) to $\gamma=1/2$ (the Gaussian statistics). Below we provide simple heuristic arguments favoring the transition in the critical exponent $\gamma(\phi)$ as a function of $\phi$.

Consider the setup depicted in \fig{fig:02}a, where the tracer particle, performing a biased random walk in $x$-direction and unbiased random walk in $y$-direction, hits the impermeable circular boundary at some angle $\phi_1$. If $\phi_1>\phi_{cr}$ (where $\phi_{cr}$ will be determined later), the tracer has ``enough'' time $t$ to run along a fish from the entry point $A$ to the destination point $B$, where ``enough'' means that the tracer equilibrates its statistics when running along the arc $AB$ (note that due to the constant force acting on the tracer in $x$-direction, the tracer's trajectory is being pressed to the curved boundary). The corresponding span, $\Delta$, which characterizes a tube within which the trajectory is localized near the boundary, has the scaling $\Delta(t)\sim t^{1/3}$.

Let us turn now to the opposite situation schematically shown in \fig{fig:02}b, which corresponds to short times $t$ (short arc $AB$). In this regime, for some angle $\phi_2<\phi_{cr}$ the arc $AB$ is so flat and so short that the tracer, hitting the boundary at point $A$, does not change its statistics when running from $A$ to $B$. Hence the fluctuations remain Gaussian (as they were before hitting the boundary) providing the scaling $\Delta(t) \sim t^{1/2}$.

To estimate $\phi_{cr}$ let us remember that the KPZ-like scaling, characterised by the dependence $\Delta \sim t^{1/3}$, emerges when the time, $t$, spent by a biased random walk when running along the arc $AB$, exceeds some longitudinal correlation time, $t_{cr} \sim R^{2/3}$. This provides the scaling dependence for the critical angle, $\phi_{cr}$, separating KPZ-like and Gaussian-like regimes,
\be
\phi_{cr}\sim \frac{t_{cr}}{R} \sim R^{-1/3}
\label{eq:03}
\ee



\section{Tracy-Widom scaling and the BBP-transition in a Jakiw-Teitelbom gravity}

In this Section we briefly discuss emergence of the Tracy-Widom scaling and the BBP-like transition in the context of Jackiw–Teitelboim (JT) gravity. We follow the logic suggested in \cite{gorsky2021lifshitz} for the holographic description of the Lifshitz tail in (1+1)-dimensional disordered system in the vicinity of the boundary. To this aim, as proposed in \cite{kitaev2019statistical,yang2019quantum,stanford2019jt}, we consider the partition function of JT gravity as the 2D partition function of a charged particle in an external imaginary transversal magnetic field. The world lines of a such a particle are closed paths of fixed length $\beta$ in the hyperbolic 2D plane in radial framing. The boundary value of the dilaton which yields the effective pressure in the polymer representation can be associated with the disorder strength in the boundary theory \cite{gorsky2021lifshitz}.

The Lagrangian of the JT gravity  involves the metric and the dilaton field, $\phi$:
\be
L_{JT}= -S_0\, \chi(M) - \int_M \phi\left(R+\frac{2}{l^2}\right) - 2\int_{\partial M}\phi\, K
\label{lag}
\ee
where $\chi(M)$ is the Euler characteristic of the manifold $M$, $l$ is the $AdS_2$ radius, and the last term is the boundary Gibbons-Hawking term. We are interested in the partition function of the JT gravity in the disc with the fixed boundary length, $L$, and fixed value of the boundary dilaton, $\phi_b = p\,l^2$:
\be
Z(L,p)= \int_{L,p} Dg_{\mu\nu} D\phi e^{-S_{JT}}
\label{eq:z}
\ee
Variation of the Lagrangian \eq{lag} with respect to the dilaton defines the radius $R$ of the two-dimensional hyperbolic disc: $R=-\frac{2}{l^2}$. The JT action reads now
\be
S_{JT}= -2\pi p\,l^2 - p\,A
\ee
where $A$ is the hyperbolic disc area. Thus, the path integral \eq{eq:z} gets reduced to the path integral over loops weighted with the enclosed area $A$ in $AdS_2$:
\be
Z(L,p)=\int \frac{D\{\rm loops\}}{SL(2,R)}\, e^{p\,A +2\pi p\, l^2}
\label{path}
\ee
The integral in \eq{path} runs over the space factorized by the action of the symmetry group $SL(2,R)$. Discretizing the path and representing the loop as a freely joint chain of $N$ segments of length $a$ each, we can express the renormalized length $\beta$ in terms of $N$, $a$ and $l$: $\beta= \frac{Na^2}{2l}$.

It was found in \cite{kitaev2019statistical,yang2019quantum} that the partition function $Z_{JT}(\beta)$ has an appealing realization as the partition function of a particle in the Euclidean time on the hyperbolic plane $H_2$ in the effective external transverse magnetic field, i.e.
\be
Z_{part}(\beta,p)= \sum_{\rm loops} e^{p\, A}
\ee
where $p$ is the chemical potential of the area $A$ in the canonical ensemble and the summation runs over all loops. The explicit relation between $Z_{JT}$ and $Z_{part}$ is as follows:
\be
Z_{JT}(\beta,p)= Z_{part}(\beta,p)
\ee
The partition function $Z_{JT}(\beta,p)$ can be considered as the Laplace transform of the spectral density, $\rho(E)$:
\be
Z_{JT}(\beta)=\int dE \rho(E) e^{-\beta E}
\ee
Various regimes in the $(\beta,p)$ plane amount to different forms of the spectral density, $\rho(E)$, which were identified in \cite{stanford}. Equivalently $\rho(E)$ has a meaning of the polymer partition function at the temperature $\beta=T^{-1}$ (see \cite{gorsky2021lifshitz} for detail). Now we can interpret both Tracy-Widom and Gauss scaling laws in terms of $\rho(E)$ in this setup. As it was shown in \cite{gorsky2021lifshitz}, using the Laplace transform of the JT gravity partition function to the spectral density in the boundary quantum mechanics, we can associate the KPZ regime with the Lifshitz tail in the corresponding spectral density $\rho_{LT}(E)$ in (1+1)D:
\be
Z(\beta)=\int dE \rho_{LT}(E) e^{-\beta E}
\ee
The key point required for Tracy-Widom scaling is the radial cutoff $R_0$ in the hyperbolic plane. The TW regime occurs when the dominant trajectories of the effective particle run close to the cutoff.

\subsection{Infrared (IR) cutoff}

The IR cutoff in the hyperbolic disc we can be introduced by hands or as the half wormhole. We are interested in the situation when the effective particle fluctuates nearby the IR cutoff radius. Here we need two parameters: the inverse temperature in the boundary theory $\beta$, and the cutoff radius $R_0$. Since the length of the trajectory is the inverse temperature, such a formulation of the problem corresponds to the microcanonical (fixed-length) ensemble. Assuming $\beta=c R_0$, we introduce the relation between the large enough boundary temperature and the IR cutoff scale. Since the whole path is located nearby the $R_0$, the effect of curved geometry can be neglected. Magnetic field does not play an essential role in this limit. Therefore we find ourselves exactly with the example considered in Section II. Varying the parameter $c$ we interpolate between KPZ-like and Gaussian fluctuations.

\subsection{Ultraviolet (UV) cutoff}

\begin{figure}[ht]
    \centering
    \includegraphics[width=.6\textwidth]{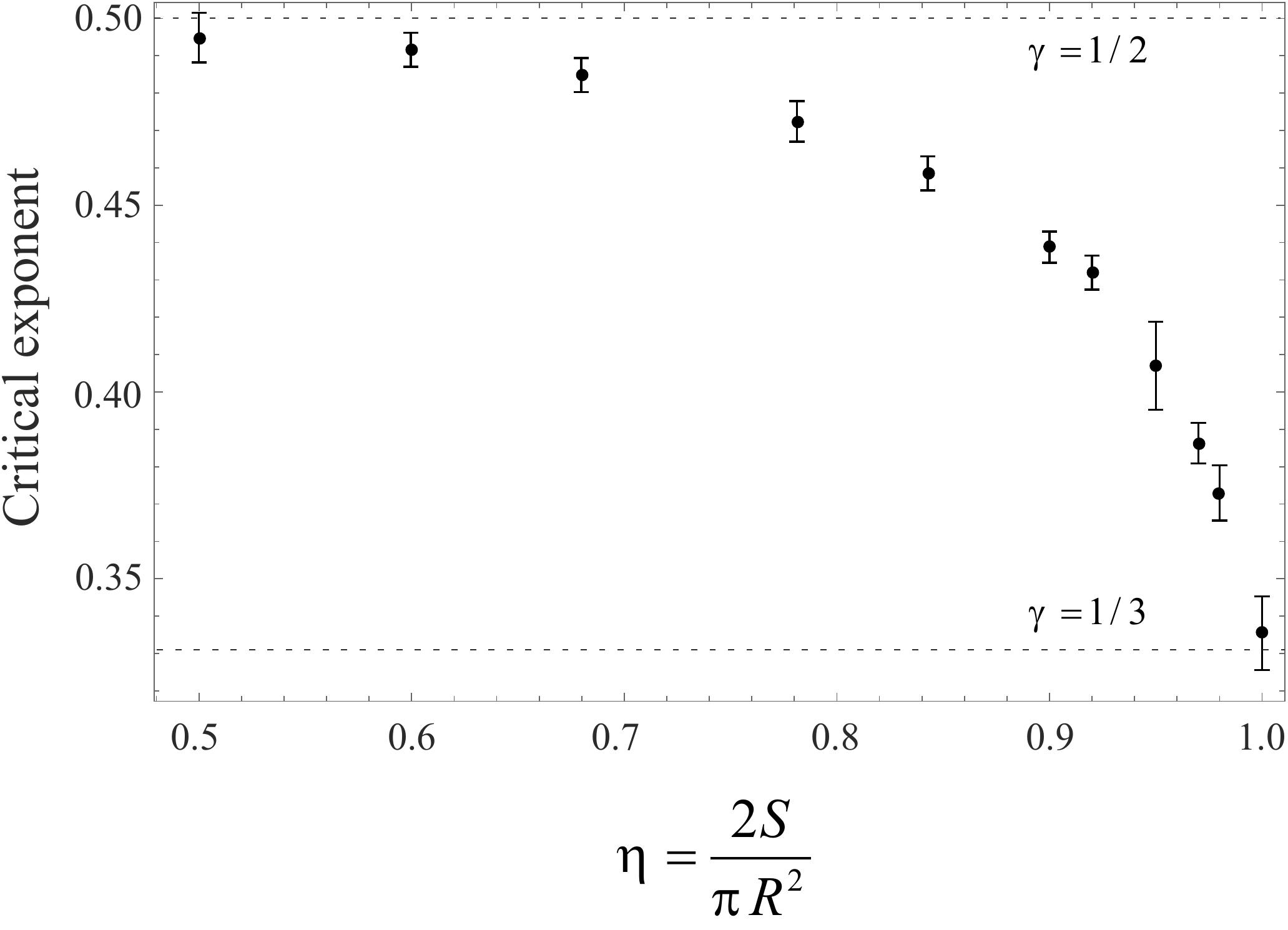}
    \caption{Behavior of the critical exponent of the inflated random loop with ``outer'' (UF) cut-off as a function of the inflation parameter $\eta=2S/(\pi R^2)$.}
    \label{edge-area}
\end{figure}

If the temperature is small enough, the typical radius of the boundary is large and we are in the Schwartzian limit. Introducing the cutoff at the UV scale, we naturally find ourselves in the mode of inflated paths of particles in the magnetic field {\it inside} a ``cavity'' of radius $R_0$. Now all three parameters matter: the magnetic field, the length of trajectory and the UV cutoff radius. We assume that the length of trajectory is slightly larger than the boundary length, $2\pi R_0$. Also we assume that the magnetic Larmour radius, $R_{Lar}$, is close to $R_0$ as well. Implying $R_{Lar}=cR_0$ and varying $\eta=2S/(\pi R^2)$, we see the transition for the KPZ to Gaussian scaling for fluctuations. The result of corresponding numerical simulations is shown in \fig{edge-area} and clearly demonstrates the transition.

Recently a possible non-perturbative completion of the JT gravity has been discussed in \cite{johnson2022distribution, johnson2022microstate}. It was based on the matrix model realization of JT gravity found in \cite{saad2019jt,stanford2019jt}. The finiteness of the entropy implies a discreteness of the spectrum in the underlying matrix model. To this aim the following specific ``quantization condition''  for the spectral density has been suggested in \cite{johnson2022distribution,johnson2022microstate}
\be
\int \rho(E) = n
\ee
where the effective cutoff in the energy is introduced. The quantization condition amounts to the formation of gaps in the spectrum of the boundary theory and ensures the finiteness of the entropy. The Fredholm determinant representation for the distribution of spectral gaps opens the door for the emergence of the Tracy-Widom distribution for fluctuations (recall that TW distribution is tightly related to the Fredholm determinants via the solutions of the Painleve II equation).

Yet we considered the representation of the JT partition function in terms of the ensemble of paths. It is eligible to ask whether the spectral view of the Tracy-Widom distribution \cite{johnson2022distribution,johnson2022microstate} has a relation to our path integral consideration. The attempt to answer this question has lead us to the following conjecture. Note that we are interested in the span, $\Delta=\sqrt{\la(r-\la r\ra)^2\ra}$ of radial fluctuations in the ensemble of paths in $AdS_2$. However, the radial direction in $AdS_2$ has the meaning of the energy scale in the boundary theory. Hence the statistics of radial fluctuations in ensemble of trajectories indeed can be viewed as a statistics of fluctuations in the energy space, $\sqrt{\la(E-\la E \ra)^2\ra}$. 

Besides, details are different. The condition implied on paths to accumulate in the vicinity of the radial cutoff is the counterpart of the claim that one focuses at the fluctuations near the cutoff in the energy space. In \cite{johnson2022distribution,johnson2022microstate} the TW was identified with the distribution of gaps in the spectrum. Qualitatively the TW distribution emerges for a very similar object, however deeper clarification of this point is required. Searching for microstates in a JT gravity representing near-horizon microstates (of near-to-extremal charged black hole) in terms of the path integral on the hyperbolic plane in an external magnetic field seems a challenging question. We postpone a more detailed discussion of this issue and the emergence of the BBP transition in JT gravity with cutoff for the forthcoming study.

\section{Conclusion}

In our work we have reformulated the Baik-Ben Arous-P\'ech\'e (BBP) transition, found earlier for the largest eigenvalue fluctuations, in terms of the scaling of fluctuations in a bunch of trajectories contributing to the free energy of a single polymer. We have found numerically in several model systems manifestations of the BBP transition from the Tracy-Widom scaling to the Gaussian one for paths' fluctuations in the (1+1)D Euclidean plane under the variation of some control parameters. Despite these control parameters could vary from one model to the other, they all have a common feature: they govern the strength of the localization of trajectories down to some region. More specifically, using these parameters we artificially force trajectories of particles (or shape of linear ideal polymers) to fluctuate nearby a solid, or partially permeable convex boundaries. Such a forced regime supports the BBP transition occurring globally along the whole trajectory. To look at the BBP transition from different perspectives, we have considered microcanonical (fixed paths length) as well as canonical ensembles.

Geometric arguments supporting the BBP transition in path integral formulation can be applied to the Jakiw-Teitelbom (JT) gravity by extending the arguments of the work \cite{gorsky2021lifshitz}. The JT gravity can be formulated for an ensemble of fixed length paths of an effective massive particle in an external magnetic field (the temperature in the boundary theory). Introducing the radial cutoff in the hyperbolic plane with the cutoff radius close to the Larmour radius of the effective particle, we find ourselves in the situation of stretched trajectories and hence have the BBP-like regime for fluctuations under the variation of the control parameter.

We believe that the model considered in our work is sufficiently rich and provides new challenging questions related to the BBP transition. Namely, we have identified numerically the BBP transition for fluctuations of paths near a semipermeable boundary. It has been mention at length of this paper that this pattern mimics fluctuations of a selected polymer in an ensemble of interacting vicious walks. For completely impermeable $(\eta=0)$ disc, we are in case of fermionic paths. At $0<\eta<1$ the polymer partially penetrates the disc and can be viewed as a path with the anyonic statistics. At $\eta=1$ the polymer does not interact with the disc at all. It would be very interesting to compare this interpretation to the anyonic Calogero model for the $\beta$-matrix ensemble and to derive the corresponding BBP transition analytically.

Another promising question concerns the identification of the BBP transition for phase-space trajectories. In this case the Planck constant is a kind of control parameter together with the energy. At the classical level we have the non-fluctuating phase space trajectories. However when the quantum corrections matter we can discuss the fluctuations of trajectories around the classical one and the TW law for these fluctuations is expected at the peculiar semiclassical approximation. 

Interesting open questions deal with understanding details of the BBP transition for the dynamical systems when it is considered via path integrals, or it is studied using its spectral dual. As example of the relevant duality the duality between the inhomogeneous TASEP and the Goldfish model found in \cite{gorsky2022dualities} could be mention. The BBP transition in the inhomogeneous TASEP has been found in \cite{barraquand2015phase} when jump rates exceed some critical values. On the other hand, jump rates in TASEP correspond to the coordinates in the Goldfish model. Hence, one could expect the BBP transition in the Goldfish model when one particle is separated from the rest at the critical distance.

\begin{acknowledgments}
We are grateful to A. Grosberg and K. Polovnikov for valuable comments. A.G. is thankful to the Basis Foundation grant No. 20-1-1-23-1.
\end{acknowledgments}

\begin{appendix}

\section{Induced $(1+1)$ false vacuum decay and paths localization}

\subsection{Semiclassical picture}

The induced false vacuum decay in $(1+1)$D provides an interesting playground to investigate the BBP transition. It has a lot in common with the setup considered in \cite{krajenbrink2021tilted}, namely the polymer statistics in the 2D Euclidean space in the background of a point-like attractive defect, or a line of attractive defects. The BBP transition occurs when some part of a polymer gets trapped at the line of defects: outside of the trapped region the polymer fluctuations are Gaussian, while inside there are of KPZ-type. To see such a transition, the trapped region should be large enough to ensure the statistics with the KPZ-like scaling.

Similarly, we can formulate the problem of a false vacuum decay induced by an external particle of mass $m$. Upon the Wick rotation to the Euclidean time, the space-time trajectory of a particle in a 2D Euclidean space is identical to a random walk (``ideal polymer''), as it follows from the standard relation between the statistical mechanics and the field theory. The Brownian white noise can now be identified with the conventional quantum white noise, while the interaction (attraction) with the defect in the classical picture is replaced now by the bouncing of external particle from a kink in the quantum picture. The attraction in the quantum case occurs due to the zero mode state of the external particle on the kink and the strength of the attraction is controlled by the mass of the external particle. Below we present the semi-quantitative arguments confirming this picture.

Remind that spontaneous vacuum decay is described in the ``thin wall approximation'' via bouncing of a particle from a solid circle (kink) in the 2D Euclidean plane. The decay probability is described by the action evaluated at the classical solution supplemented by the quantum determinant \cite{voloshin1974bubbles,callan1977fate,voloshin1986particle}. To have an example, consider the $\phi^4$-scalar theory  with the potential 
\be
V(\phi) = \lambda(\phi^2 -\eta^2)^2 +\epsilon \phi
\label{eq:phi}
\ee 
where $\lambda, \eta$ and $\epsilon$ are parameters of the potential. The effective Lagrangian in leading exponential approximation \cite{voloshin1974bubbles,coleman1977fate} can be written in the radial framing as follows
\be
L_{eff}=2\pi R \mu - \pi \epsilon R^2
\ee
where $\mu$ is the kink mass, and the radial symmetry is assumed. The action evaluated at the critical bounce provides the spontaneous decay probability $\omega$ in the leading approximation:  
\be
\omega \propto e^{-\frac{\mu^2}{\epsilon}}
\label{eq:omega}
\ee
The pre-exponential factor for the decay $\omega$ in \eq{eq:omega} can be evaluated as well \cite{kiselev1984calculation} in the thin wall approximation. The general expression for the decay probability $\omega^*$ up to the exponentially small corrections is
\be
\omega^*=\frac{\epsilon}{2\pi} e^{-\frac{\mu^2}{\epsilon}}
\label{eq:omega1}
\ee
The pre-exponential factor in the thin wall approximation originates from the summation over closed paths of charged particle in the plane in a constant magnetic field $\epsilon$ with the following effective Lagrangian
\be
L=\int_{0}^{2\pi} ds\left(\mu \sqrt{{\bf r}^2(s) + \dot{{\bf r}}^2(s)} -\frac{1}{2}\epsilon {\bf r}^2(s)\right)
\ee
with $0\le s\le 2\pi$.

The induced false vacuum decay at the semiclassical level has been considered in \cite{voloshin1986particle,voloshin1994catalyzed,gorsky2006particle}. The key point is the zero-mode classical solution of binding of external particle at the kink configuration whose trajectory in the Euclidean space-time represents the bouncing solution. If the external particle is a boson, the zero mode follows from the translational invariance, while for the external fermion it follows from the index theorem. If the mass of the external particle, $m$, is much smaller than the mass of the kink, i.e. $m \ll \mu$, then the circle representing the kink is not perturbed. To the contrary, if masses of the external particle and the kink are comparable, i.e. $m \propto \mu$, the circle gets deformed into the ``fish'' similar to the one shown in \fig{fig:02} and the decay probability in the leading approximation can be written as \cite{voloshin1986particle}
\be
w \propto e^{-S}, \qquad S = \frac{2\mu^2}{\epsilon} \arcsin \frac{m}{2\mu} +
\frac{m\mu}{\epsilon}\sqrt{1-\frac{m^2}{4\mu^2}}
\label{eq:w}
\ee
Note that the considered process has two complementary interpretations. From the ``false vacuum decay'' viewpoint we deal with the decay in the presence of the external particle and evaluate the fluctuations on the top of the classical bouncing solution. However, from the ``particle'' viewpoint we evaluate the corresponding Green function in the Euclidean time, $T$, $\la x=0\left|e^{-HT}\right|x=0 \ra$. The imaginary part of the effective mass in the Green function determines the decay rate.

\subsection{Towards the BBP transition}

Now we are in position to discuss the connection between the induced false vacuum decay and the BBP transition. The setups of our model and of the one considered in \cite{krajenbrink2021tilted} are similar with only one distinction: in our case we consider a circular extended attracting defect, while in \cite{krajenbrink2021tilted} the line of point-like defects is introduced. To discuss the BBP transition we have to estimate quantum fluctuations on top of the classical solution.

It has been pointed out above that in the limit $m \gg \mu$ the external particle does not deform the defect (the circular boundary). The fluctuations outside the circular boundary are Gaussian, however the fluctuations of the trapped part of path are not. It was shown in \cite{gorsky2006particle} that at the small mass of an external particle the IR divergences cancel which means that our system is the IR-safe object. As it was discussed in a previous section, the variance, $\sqrt{\la h^2 -\la h\ra ^2\ra}$, of a typical particle-bounce distance is determined by a KPZ $1/3$ exponent. From the viewpoint of paths statistics our ensemble is canonical and each path carries the weight $e^{-mL}$, where $L$ is the paths' length. The BBP transition occurs in the vicinity of the trapping point for the external particle.

If the mass of the external particle is sufficiently large, its back reaction on the bounce configuration should be taken into account leading to the ``fish-like'' shape. The ``fish'' consists of two arcs of same radii, and the cusp angle, $\psi$, is fixed by the equilibrium condition at the vertex. This configuration is exactly as the one discussed in the previous section and is depicted in \fig{fig:02} with $\psi=2\phi$. Recall that at a particular value of the cusp angle, $\psi$, the BBP transition indeed occurs. Hence, in this case we have again the BBP transition for fluctuations of the quantum size of emerging bound state. The ratio $\xi=\frac{m}{\mu}$, where $m$ and $\mu$ are the masses of an external particle and of a kink, plays the role of the control parameter: when $\xi$ is large enough, the arcs are small, there is no enough time to form the KPZ regime, and there is no transition in this case.

\section{Nonlinear dimensional reduction for ensemble of paths}

Here we apply the dimensional reduction approach for the investigation of the BBP transition in the microcanonical ensemble of trajectories nearby the partially permeable semicircle. The BBP transition is manifested in two-dimensional manifold as a specific flow of data enveloping shape. Each point on the plane represents the particular trajectory.

The characterization of transition by scaling exponents is rather crude since it does not reflect the information about the particular structure of trajectories. Thus, in order to see the BBP transition at the level of paths, rather than their fluctuation exponents $\gamma$, the nonlinear dimension reduction algorithm UMAP \cite{mcinnes2018umap} is used. As the dataset for which dimensional reduction is performed, we choose the ensemble of 400-component vectors, ${\bf X}=(x_1,y_1,\ldots,x_N,y_N)$, uniquely encoding trajectories of length $N=200$, wandering near the disc of radius $R=50$. Our dataset uniformly covers the interval $0\leq\eta<1$ with the parameter set $\eta=\{0,0.1,0.2,...,0.9\}$, where 1400 vectors ${\bf X}$ are generated for each value $\eta$. 

\begin{figure}[H]
\centering
\includegraphics[width=.85\textwidth]{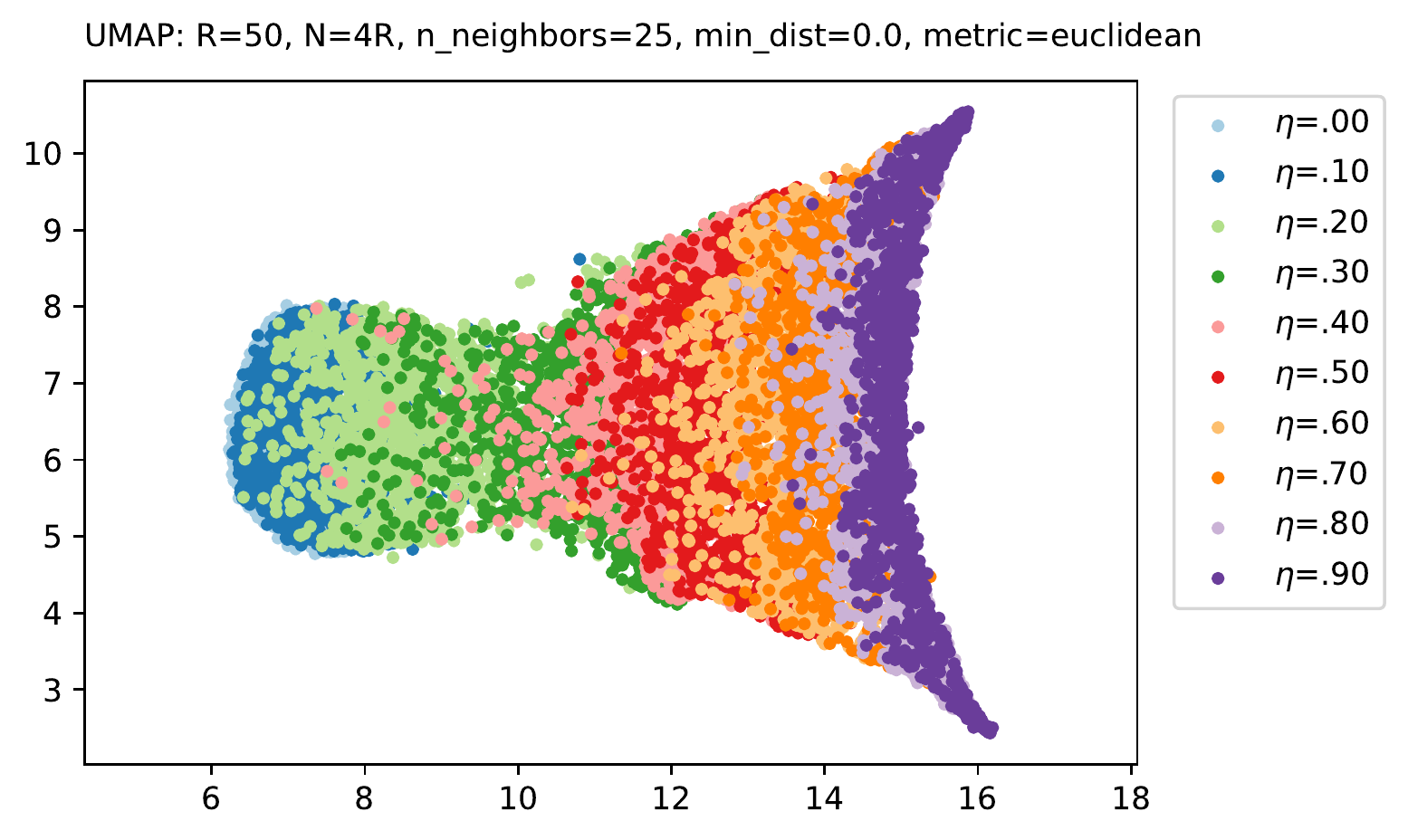}
\caption{UMAP visualization of the trajectories dataset, with embedded points are color-coded according to parameter $\eta$. }
\label{fig:03}
\end{figure}

\begin{figure}[H]
\centering
\includegraphics[width=.85\textwidth]{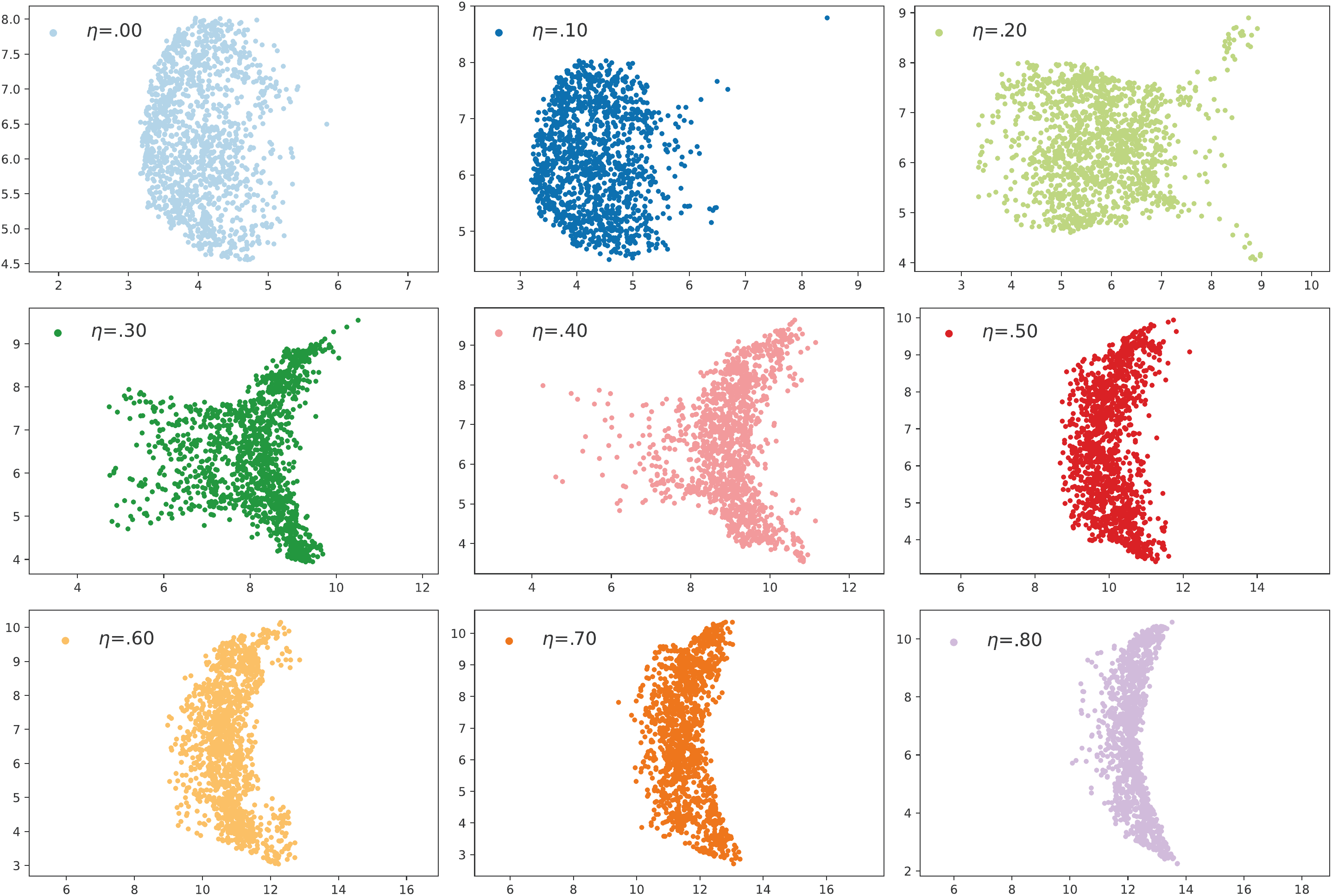}
\caption{UMAP visualization of the trajectories dataset, with embedded points are color-coded according to parameter $\eta$. }
\label{fig:04}
\end{figure}

The UMAP projection of the 400-component vectors, ${\bf X}$ to the two-dimensional phase space are shown in \fig{fig:03} where all embedded vectors are plotted simultaneously, and in \fig{fig:04}, where these vectors are plotted separately. When $\eta$ is increasing from $\eta=0$ towards $\eta=1$, the number of trajectories with ``1/3'' (KPZ) statistics is decreasing and simultaneously, the number of paths with ``1/2'' (Gaussian) statistics is increasing. Since each point in \fig{fig:03}, \fig{fig:04} designates some trajectory, we may conclude that the right spot corresponds to paths with the Gaussian statistics $\eta \in[0.5,0.9]$, while the left spot consists of paths with the KPZ statistics $\eta \in[0.0,0.2]$. We claim that the transition region $\eta \in[0.3,0.4]$ with mixed statistics corresponds to the BBP transition, which is in a qualitative agreement with the behavior seen in \fig{fig:01}. To summarize, we have demonstrated that the BBP transition upon the dimensional reduction is manifested in the formation of a ``bottleneck'' in the enveloping shape of data flow on a two-dimensional plane upon changing the control parameter.

\section{BBP transition for distribution of largest eigenvalue}

Let us briefly recall for completeness the standard formulation of the BBP transition in terms of the distribution of the largest eigenvalues of certain non-null complex $N\times N$ positive covariance matrices $S$ \cite{baik2005phase,baik2006painleve}. Consider $M$ vectors $({\bf y}_1\dots {\bf y}_M)$ each of dimension $N$. The density of all vectors is Gaussian with the mean value ${\bf \mu}$:
\be
P({\bf y})= \frac{1}{(2\pi)^{N/2}(\det\Sigma)^{1/2}} \exp\left(-\frac{1}{2}\left< {\bf y}-{\bf \mu},\Sigma^{-1},
{\bf y}-{\bf \mu}\right>\right)
\ee

Construct now the $N\times M$ matrix $X=\left[{\bf y}_1 - {\bf Y}, \dots {\bf y}_M - {\bf Y} \right]$ and combine it into  $N\times N$ covariance matrix $S$ as follows: $S=XX^{t}$. Assume that $M\to \infty$ and $N\to \infty$ such that their ratio is finite $M/N=\varkappa$. The matrix $S$ belongs to the Laguerre or Wishart ensemble whose asymptotic properties at $\Sigma={\rm Id}$ are known. The corresponding spectral density obeys the Marchenko-Pastur law \cite{marchenko1967distribution}
\be
\rho(x)= \frac{\varkappa^2}{2x}\sqrt{(b-x)(x-a)}, \qquad a<x<b 
\ee
where
\be
a=\left(\frac{\varkappa-1}{\varkappa}\right)^2; \qquad b=\left(\frac{\varkappa+1}{\varkappa}\right)^2
\ee
The largest eigenvalue fixing the spectral edge is $\lambda_{max}=b$ and the spectral fluctuations at the spectral edge obey the Tracy-Widom distribution
\be
P\left((\lambda_{max} - b) \frac{\varkappa}{(1+\varkappa)^{4/3}} M^{2/3}\leq s\right) \rightarrow  F_{GOE}(s)
\ee
where $F_{GOE}$ is the TW distribution for orthogonal ensemble. Now departure from the identity matrix is introduced and some number of degenerate non-unit eigenvalues of the covariance matrix $\Sigma$ are selected $l_1 \dots l_k\neq 1$. The value $l_1$ is considered as the control parameter for the BBP phase transition.

For the complex Gaussian samples \cite{baik2005phase} the critical value of the non-unit covariance eigenvalue for perturbed case is $l_{crit}= 1 +\varkappa^{-1}$, where $\gamma$ is parameter of the ensemble.
\be
P\left((\lambda_{max}-l_{crit}^2) \frac{\varkappa}{(1+\varkappa)^{4/3}} N^{2/3}\leq s\right) \rightarrow  F_k(s)
\ee
The function $F_k(s)$ is the eigenvalue distribution at the transition point, where the eigenvalue $l_1$ of the covariance matrix has the multiplicity $k$ and is equal to the critical value. If $k$ non-unit eigenvalues of the covariance matrix are equal and lie above the critical value, $l_{crit}$, i.e. $l_i> l_{crit}$, the Gaussian unitary ensemble distribution, $G_k(s)$
enters the game
\be
P\left(\left(\lambda_{max} - \left(l_1 + \frac{l_1 \varkappa^{-2}}{l_1-1}\right)\right) N^{1/2}\sqrt{l_1^2 -  \frac{l_1^2 \varkappa^{-2}}{(l_1-1)^2}} \leq s\right) \rightarrow  G_k(s)
\ee

\end{appendix}
\bibliography{references.bib}

\begin{thebibliography}{43}
\expandafter\ifx\csname natexlab\endcsname\relax\def\natexlab#1{#1}\fi
\expandafter\ifx\csname bibnamefont\endcsname\relax
  \def\bibnamefont#1{#1}\fi
\expandafter\ifx\csname bibfnamefont\endcsname\relax
  \def\bibfnamefont#1{#1}\fi
\expandafter\ifx\csname citenamefont\endcsname\relax
  \def\citenamefont#1{#1}\fi
\expandafter\ifx\csname url\endcsname\relax
  \def\url#1{\texttt{#1}}\fi
\expandafter\ifx\csname urlprefix\endcsname\relax\def\urlprefix{URL }\fi
\providecommand{\bibinfo}[2]{#2}
\providecommand{\eprint}[2][]{\url{#2}}

\bibitem[{\citenamefont{Tracy and Widom}(2009)}]{tracy2009distributions}
\bibinfo{author}{\bibfnamefont{C.~A.} \bibnamefont{Tracy}} \bibnamefont{and}
  \bibinfo{author}{\bibfnamefont{H.}~\bibnamefont{Widom}},
  \bibinfo{journal}{New trends in mathematical physics} pp.
  \bibinfo{pages}{753--765} (\bibinfo{year}{2009}).

\bibitem[{\citenamefont{Baik et~al.}(2005)\citenamefont{Baik, Arous, and
  P{\'e}ch{\'e}}}]{baik2005phase}
\bibinfo{author}{\bibfnamefont{J.}~\bibnamefont{Baik}},
  \bibinfo{author}{\bibfnamefont{G.~B.} \bibnamefont{Arous}}, \bibnamefont{and}
  \bibinfo{author}{\bibfnamefont{S.}~\bibnamefont{P{\'e}ch{\'e}}},
  \bibinfo{journal}{The Annals of Probability} \textbf{\bibinfo{volume}{33}},
  \bibinfo{pages}{1643} (\bibinfo{year}{2005}).

\bibitem[{\citenamefont{Kosterlitz et~al.}(1976)\citenamefont{Kosterlitz,
  Thouless, and Jones}}]{kosterlitz1976spherical}
\bibinfo{author}{\bibfnamefont{J.~M.} \bibnamefont{Kosterlitz}},
  \bibinfo{author}{\bibfnamefont{D.~J.} \bibnamefont{Thouless}},
  \bibnamefont{and} \bibinfo{author}{\bibfnamefont{R.~C.} \bibnamefont{Jones}},
  \bibinfo{journal}{Physical Review Letters} \textbf{\bibinfo{volume}{36}},
  \bibinfo{pages}{1217} (\bibinfo{year}{1976}).

\bibitem[{\citenamefont{Baik}(2006)}]{baik2006painleve}
\bibinfo{author}{\bibfnamefont{J.}~\bibnamefont{Baik}}, \bibinfo{journal}{Duke
  Mathematical Journal} \textbf{\bibinfo{volume}{133}}, \bibinfo{pages}{205}
  (\bibinfo{year}{2006}).

\bibitem[{\citenamefont{Bassler et~al.}(2009)\citenamefont{Bassler, Forrester,
  and Frankel}}]{bassler2009eigenvalue}
\bibinfo{author}{\bibfnamefont{K.~E.} \bibnamefont{Bassler}},
  \bibinfo{author}{\bibfnamefont{P.~J.} \bibnamefont{Forrester}},
  \bibnamefont{and} \bibinfo{author}{\bibfnamefont{N.~E.}
  \bibnamefont{Frankel}}, \bibinfo{journal}{Journal of mathematical physics}
  \textbf{\bibinfo{volume}{50}}, \bibinfo{pages}{033302}
  (\bibinfo{year}{2009}).

\bibitem[{\citenamefont{Baik et~al.}(2018)\citenamefont{Baik, Barraquand,
  Corwin, and Suidan}}]{lpp}
\bibinfo{author}{\bibfnamefont{J.}~\bibnamefont{Baik}},
  \bibinfo{author}{\bibfnamefont{G.}~\bibnamefont{Barraquand}},
  \bibinfo{author}{\bibfnamefont{I.}~\bibnamefont{Corwin}}, \bibnamefont{and}
  \bibinfo{author}{\bibfnamefont{T.}~\bibnamefont{Suidan}},
  \bibinfo{journal}{The Annals of Probability} \textbf{\bibinfo{volume}{46}},
  \bibinfo{pages}{3015 } (\bibinfo{year}{2018}),
  \urlprefix\url{https://doi.org/10.1214/17-AOP1226}.

\bibitem[{\citenamefont{Borodin et~al.}(2014)\citenamefont{Borodin, Corwin, and
  Ferrari}}]{borodin2014free}
\bibinfo{author}{\bibfnamefont{A.}~\bibnamefont{Borodin}},
  \bibinfo{author}{\bibfnamefont{I.}~\bibnamefont{Corwin}}, \bibnamefont{and}
  \bibinfo{author}{\bibfnamefont{P.}~\bibnamefont{Ferrari}},
  \bibinfo{journal}{Communications on Pure and Applied Mathematics}
  \textbf{\bibinfo{volume}{67}}, \bibinfo{pages}{1129} (\bibinfo{year}{2014}).

\bibitem[{\citenamefont{Aggarwal and Borodin}(2019)}]{aggarwal2019phase}
\bibinfo{author}{\bibfnamefont{A.}~\bibnamefont{Aggarwal}} \bibnamefont{and}
  \bibinfo{author}{\bibfnamefont{A.}~\bibnamefont{Borodin}},
  \bibinfo{journal}{The Annals of Probability} \textbf{\bibinfo{volume}{47}},
  \bibinfo{pages}{613} (\bibinfo{year}{2019}).

\bibitem[{\citenamefont{Barraquand}(2015)}]{barraquand2015phase}
\bibinfo{author}{\bibfnamefont{G.}~\bibnamefont{Barraquand}},
  \bibinfo{journal}{Stochastic Processes and their Applications}
  \textbf{\bibinfo{volume}{125}}, \bibinfo{pages}{2674} (\bibinfo{year}{2015}).

\bibitem[{\citenamefont{Saber and Saberi}(2022)}]{saber2022universal}
\bibinfo{author}{\bibfnamefont{S.}~\bibnamefont{Saber}} \bibnamefont{and}
  \bibinfo{author}{\bibfnamefont{A.~A.} \bibnamefont{Saberi}},
  \bibinfo{journal}{Physical Review E} \textbf{\bibinfo{volume}{105}},
  \bibinfo{pages}{L022102} (\bibinfo{year}{2022}).

\bibitem[{\citenamefont{Baik and Lee}(2020)}]{glass}
\bibinfo{author}{\bibfnamefont{J.}~\bibnamefont{Baik}} \bibnamefont{and}
  \bibinfo{author}{\bibfnamefont{J.~O.} \bibnamefont{Lee}},
  \bibinfo{journal}{Annales de l'Institut Henri Poincaré, Probabilités et
  Statistiques} \textbf{\bibinfo{volume}{56}}, \bibinfo{pages}{2897 }
  (\bibinfo{year}{2020}), \urlprefix\url{https://doi.org/10.1214/20-AIHP1062}.

\bibitem[{\citenamefont{Adler et~al.}(2009)\citenamefont{Adler, Del{\'e}pine,
  and Van~Moerbeke}}]{adler2009dyson}
\bibinfo{author}{\bibfnamefont{M.}~\bibnamefont{Adler}},
  \bibinfo{author}{\bibfnamefont{J.}~\bibnamefont{Del{\'e}pine}},
  \bibnamefont{and}
  \bibinfo{author}{\bibfnamefont{P.}~\bibnamefont{Van~Moerbeke}},
  \bibinfo{journal}{Communications on Pure and Applied Mathematics: A Journal
  Issued by the Courant Institute of Mathematical Sciences}
  \textbf{\bibinfo{volume}{62}}, \bibinfo{pages}{334} (\bibinfo{year}{2009}).

\bibitem[{\citenamefont{Krajenbrink et~al.}(2021)\citenamefont{Krajenbrink,
  Le~Doussal, and O'Connell}}]{krajenbrink2021tilted}
\bibinfo{author}{\bibfnamefont{A.}~\bibnamefont{Krajenbrink}},
  \bibinfo{author}{\bibfnamefont{P.}~\bibnamefont{Le~Doussal}},
  \bibnamefont{and}
  \bibinfo{author}{\bibfnamefont{N.}~\bibnamefont{O'Connell}},
  \bibinfo{journal}{Physical Review E} \textbf{\bibinfo{volume}{103}},
  \bibinfo{pages}{042120} (\bibinfo{year}{2021}).

\bibitem[{\citenamefont{Nechaev et~al.}(2019)\citenamefont{Nechaev, Polovnikov,
  Shlosman, Valov, and Vladimirov}}]{nechaev2019anomalous}
\bibinfo{author}{\bibfnamefont{S.}~\bibnamefont{Nechaev}},
  \bibinfo{author}{\bibfnamefont{K.}~\bibnamefont{Polovnikov}},
  \bibinfo{author}{\bibfnamefont{S.}~\bibnamefont{Shlosman}},
  \bibinfo{author}{\bibfnamefont{A.}~\bibnamefont{Valov}}, \bibnamefont{and}
  \bibinfo{author}{\bibfnamefont{A.}~\bibnamefont{Vladimirov}},
  \bibinfo{journal}{Physical Review E} \textbf{\bibinfo{volume}{99}},
  \bibinfo{pages}{012110} (\bibinfo{year}{2019}).

\bibitem[{\citenamefont{Vladimirov et~al.}(2020)\citenamefont{Vladimirov,
  Shlosman, and Nechaev}}]{vladimirov2020brownian}
\bibinfo{author}{\bibfnamefont{A.}~\bibnamefont{Vladimirov}},
  \bibinfo{author}{\bibfnamefont{S.}~\bibnamefont{Shlosman}}, \bibnamefont{and}
  \bibinfo{author}{\bibfnamefont{S.}~\bibnamefont{Nechaev}},
  \bibinfo{journal}{Physical Review E} \textbf{\bibinfo{volume}{102}},
  \bibinfo{pages}{012124} (\bibinfo{year}{2020}).

\bibitem[{\citenamefont{Gorsky et~al.}(2018)\citenamefont{Gorsky, Nechaev, and
  Valov}}]{gorsky2018statistical}
\bibinfo{author}{\bibfnamefont{A.}~\bibnamefont{Gorsky}},
  \bibinfo{author}{\bibfnamefont{S.}~\bibnamefont{Nechaev}}, \bibnamefont{and}
  \bibinfo{author}{\bibfnamefont{A.}~\bibnamefont{Valov}},
  \bibinfo{journal}{Journal of High Energy Physics}
  \textbf{\bibinfo{volume}{2018}}, \bibinfo{pages}{1} (\bibinfo{year}{2018}).

\bibitem[{\citenamefont{Valov et~al.}(2021)\citenamefont{Valov, Gorsky, and
  Nechaev}}]{valov2021equilibrium}
\bibinfo{author}{\bibfnamefont{A.}~\bibnamefont{Valov}},
  \bibinfo{author}{\bibfnamefont{A.}~\bibnamefont{Gorsky}}, \bibnamefont{and}
  \bibinfo{author}{\bibfnamefont{S.}~\bibnamefont{Nechaev}},
  \bibinfo{journal}{Physics of Particles and Nuclei}
  \textbf{\bibinfo{volume}{52}}, \bibinfo{pages}{185} (\bibinfo{year}{2021}).

\bibitem[{\citenamefont{Ferrari and Spohn}(2005)}]{ferrari2005constrained}
\bibinfo{author}{\bibfnamefont{P.~L.} \bibnamefont{Ferrari}} \bibnamefont{and}
  \bibinfo{author}{\bibfnamefont{H.}~\bibnamefont{Spohn}},
  \bibinfo{journal}{The Annals of Probability} \textbf{\bibinfo{volume}{33}},
  \bibinfo{pages}{1302} (\bibinfo{year}{2005}).

\bibitem[{\citenamefont{Meerson and Smith}(2019)}]{meerson2019geometrical}
\bibinfo{author}{\bibfnamefont{B.}~\bibnamefont{Meerson}} \bibnamefont{and}
  \bibinfo{author}{\bibfnamefont{N.~R.} \bibnamefont{Smith}},
  \bibinfo{journal}{Journal of Physics A: Mathematical and Theoretical}
  \textbf{\bibinfo{volume}{52}}, \bibinfo{pages}{415001}
  (\bibinfo{year}{2019}).

\bibitem[{\citenamefont{Smith and Meerson}(2019)}]{smith2019geometrical}
\bibinfo{author}{\bibfnamefont{N.~R.} \bibnamefont{Smith}} \bibnamefont{and}
  \bibinfo{author}{\bibfnamefont{B.}~\bibnamefont{Meerson}},
  \bibinfo{journal}{Journal of Statistical Mechanics: Theory and Experiment}
  \textbf{\bibinfo{volume}{2019}}, \bibinfo{pages}{023205}
  (\bibinfo{year}{2019}).

\bibitem[{\citenamefont{Kitaev and Suh}(2019)}]{kitaev2019statistical}
\bibinfo{author}{\bibfnamefont{A.}~\bibnamefont{Kitaev}} \bibnamefont{and}
  \bibinfo{author}{\bibfnamefont{S.~J.} \bibnamefont{Suh}},
  \bibinfo{journal}{Journal of High Energy Physics}
  \textbf{\bibinfo{volume}{2019}}, \bibinfo{pages}{1} (\bibinfo{year}{2019}).

\bibitem[{\citenamefont{Yang}(2019)}]{yang2019quantum}
\bibinfo{author}{\bibfnamefont{Z.}~\bibnamefont{Yang}},
  \bibinfo{journal}{Journal of High Energy Physics}
  \textbf{\bibinfo{volume}{2019}}, \bibinfo{pages}{1} (\bibinfo{year}{2019}).

\bibitem[{\citenamefont{Dotsenko}(2010)}]{dotsenko}
\bibinfo{author}{\bibfnamefont{V.}~\bibnamefont{Dotsenko}},
  \bibinfo{journal}{{EPL} (Europhysics Letters)} \textbf{\bibinfo{volume}{90}},
  \bibinfo{pages}{20003} (\bibinfo{year}{2010}),
  \urlprefix\url{https://doi.org/10.1209/0295-5075/90/20003}.

\bibitem[{\citenamefont{Calabrese and Le~Doussal}(2011)}]{doussal-kpz}
\bibinfo{author}{\bibfnamefont{P.}~\bibnamefont{Calabrese}} \bibnamefont{and}
  \bibinfo{author}{\bibfnamefont{P.}~\bibnamefont{Le~Doussal}},
  \bibinfo{journal}{Phys. Rev. Lett.} \textbf{\bibinfo{volume}{106}},
  \bibinfo{pages}{250603} (\bibinfo{year}{2011}),
  \urlprefix\url{https://link.aps.org/doi/10.1103/PhysRevLett.106.250603}.

\bibitem[{\citenamefont{Gorsky et~al.}(2021)\citenamefont{Gorsky, Nechaev, and
  Valov}}]{gorsky2021lifshitz}
\bibinfo{author}{\bibfnamefont{A.}~\bibnamefont{Gorsky}},
  \bibinfo{author}{\bibfnamefont{S.}~\bibnamefont{Nechaev}}, \bibnamefont{and}
  \bibinfo{author}{\bibfnamefont{A.}~\bibnamefont{Valov}},
  \bibinfo{journal}{Journal of High Energy Physics}
  \textbf{\bibinfo{volume}{2021}}, \bibinfo{pages}{1} (\bibinfo{year}{2021}).

\bibitem[{\citenamefont{Majumdar and Schehr}(2014)}]{majumdar2014top}
\bibinfo{author}{\bibfnamefont{S.~N.} \bibnamefont{Majumdar}} \bibnamefont{and}
  \bibinfo{author}{\bibfnamefont{G.}~\bibnamefont{Schehr}},
  \bibinfo{journal}{Journal of Statistical Mechanics: Theory and Experiment}
  \textbf{\bibinfo{volume}{2014}}, \bibinfo{pages}{P01012}
  (\bibinfo{year}{2014}).

\bibitem[{\citenamefont{Polovnikov et~al.}(2022)\citenamefont{Polovnikov,
  Nechaev, and Grosberg}}]{grosberg}
\bibinfo{author}{\bibfnamefont{K.~E.} \bibnamefont{Polovnikov}},
  \bibinfo{author}{\bibfnamefont{S.~K.} \bibnamefont{Nechaev}},
  \bibnamefont{and} \bibinfo{author}{\bibfnamefont{A.~Y.}
  \bibnamefont{Grosberg}}, \bibinfo{journal}{Phys. Rev. Lett.}
  \textbf{\bibinfo{volume}{129}}, \bibinfo{pages}{097801}
  (\bibinfo{year}{2022}),
  \urlprefix\url{https://link.aps.org/doi/10.1103/PhysRevLett.129.097801}.

\bibitem[{\citenamefont{Wilson and Poon}(2011)}]{poon}
\bibinfo{author}{\bibfnamefont{L.~G.} \bibnamefont{Wilson}} \bibnamefont{and}
  \bibinfo{author}{\bibfnamefont{W.~C.~K.} \bibnamefont{Poon}},
  \bibinfo{journal}{Phys. Chem. Chem. Phys.} \textbf{\bibinfo{volume}{13}},
  \bibinfo{pages}{10617} (\bibinfo{year}{2011}),
  \urlprefix\url{http://dx.doi.org/10.1039/C0CP01564D}.

\bibitem[{\citenamefont{Stanford and Witten}(2019)}]{stanford2019jt}
\bibinfo{author}{\bibfnamefont{D.}~\bibnamefont{Stanford}} \bibnamefont{and}
  \bibinfo{author}{\bibfnamefont{E.}~\bibnamefont{Witten}},
  \bibinfo{journal}{arXiv preprint arXiv:1907.03363}  (\bibinfo{year}{2019}).

\bibitem[{\citenamefont{Stanford and Yang}(2020)}]{stanford}
\bibinfo{author}{\bibfnamefont{D.}~\bibnamefont{Stanford}} \bibnamefont{and}
  \bibinfo{author}{\bibfnamefont{Z.}~\bibnamefont{Yang}}
  (\bibinfo{year}{2020}), \urlprefix\url{https://arxiv.org/abs/2004.08005}.

\bibitem[{\citenamefont{Johnson}(2022{\natexlab{a}})}]{johnson2022distribution}
\bibinfo{author}{\bibfnamefont{C.~V.} \bibnamefont{Johnson}},
  \bibinfo{journal}{arXiv preprint arXiv:2206.00692}
  (\bibinfo{year}{2022}{\natexlab{a}}).

\bibitem[{\citenamefont{Johnson}(2022{\natexlab{b}})}]{johnson2022microstate}
\bibinfo{author}{\bibfnamefont{C.~V.} \bibnamefont{Johnson}},
  \bibinfo{journal}{arXiv preprint arXiv:2201.11942}
  (\bibinfo{year}{2022}{\natexlab{b}}).

\bibitem[{\citenamefont{Saad et~al.}(2019)\citenamefont{Saad, Shenker, and
  Stanford}}]{saad2019jt}
\bibinfo{author}{\bibfnamefont{P.}~\bibnamefont{Saad}},
  \bibinfo{author}{\bibfnamefont{S.~H.} \bibnamefont{Shenker}},
  \bibnamefont{and} \bibinfo{author}{\bibfnamefont{D.}~\bibnamefont{Stanford}},
  \bibinfo{journal}{arXiv preprint arXiv:1903.11115}  (\bibinfo{year}{2019}).

\bibitem[{\citenamefont{Gorsky et~al.}(2022)\citenamefont{Gorsky, Vasilyev, and
  Zotov}}]{gorsky2022dualities}
\bibinfo{author}{\bibfnamefont{A.}~\bibnamefont{Gorsky}},
  \bibinfo{author}{\bibfnamefont{M.}~\bibnamefont{Vasilyev}}, \bibnamefont{and}
  \bibinfo{author}{\bibfnamefont{A.}~\bibnamefont{Zotov}},
  \bibinfo{journal}{Journal of High Energy Physics}
  \textbf{\bibinfo{volume}{2022}}, \bibinfo{pages}{1} (\bibinfo{year}{2022}).

\bibitem[{\citenamefont{Voloshin et~al.}(1974)\citenamefont{Voloshin, Kobzarev,
  and Okun}}]{voloshin1974bubbles}
\bibinfo{author}{\bibfnamefont{M.}~\bibnamefont{Voloshin}},
  \bibinfo{author}{\bibfnamefont{I.~Y.} \bibnamefont{Kobzarev}},
  \bibnamefont{and} \bibinfo{author}{\bibfnamefont{L.~B.} \bibnamefont{Okun}},
  \bibinfo{journal}{Yad. Fiz.} \textbf{\bibinfo{volume}{20}},
  \bibinfo{pages}{1229} (\bibinfo{year}{1974}).

\bibitem[{\citenamefont{Callan~Jr and Coleman}(1977)}]{callan1977fate}
\bibinfo{author}{\bibfnamefont{C.~G.} \bibnamefont{Callan~Jr}}
  \bibnamefont{and} \bibinfo{author}{\bibfnamefont{S.}~\bibnamefont{Coleman}},
  \bibinfo{journal}{Physical Review D} \textbf{\bibinfo{volume}{16}},
  \bibinfo{pages}{1762} (\bibinfo{year}{1977}).

\bibitem[{\citenamefont{Voloshin and Selivanov}(1986)}]{voloshin1986particle}
\bibinfo{author}{\bibfnamefont{M.}~\bibnamefont{Voloshin}} \bibnamefont{and}
  \bibinfo{author}{\bibfnamefont{K.}~\bibnamefont{Selivanov}},
  \bibinfo{journal}{Sov. J. Nucl. Phys.(Engl. Transl.);(United States)}
  \textbf{\bibinfo{volume}{44}} (\bibinfo{year}{1986}).

\bibitem[{\citenamefont{Coleman}(1977)}]{coleman1977fate}
\bibinfo{author}{\bibfnamefont{S.}~\bibnamefont{Coleman}},
  \bibinfo{journal}{Physical Review D} \textbf{\bibinfo{volume}{15}},
  \bibinfo{pages}{2929} (\bibinfo{year}{1977}).

\bibitem[{\citenamefont{Kiselev and Selivanov}(1984)}]{kiselev1984calculation}
\bibinfo{author}{\bibfnamefont{V.}~\bibnamefont{Kiselev}} \bibnamefont{and}
  \bibinfo{author}{\bibfnamefont{K.}~\bibnamefont{Selivanov}},
  \bibinfo{journal}{Pis' ma v Zhurnal Ehksperimental'noj i Teoreticheskoj
  Fiziki} \textbf{\bibinfo{volume}{39}}, \bibinfo{pages}{72}
  (\bibinfo{year}{1984}).

\bibitem[{\citenamefont{Voloshin}(1994)}]{voloshin1994catalyzed}
\bibinfo{author}{\bibfnamefont{M.}~\bibnamefont{Voloshin}},
  \bibinfo{journal}{Physical Review D} \textbf{\bibinfo{volume}{49}},
  \bibinfo{pages}{2014} (\bibinfo{year}{1994}).

\bibitem[{\citenamefont{Gorsky and Voloshin}(2006)}]{gorsky2006particle}
\bibinfo{author}{\bibfnamefont{A.}~\bibnamefont{Gorsky}} \bibnamefont{and}
  \bibinfo{author}{\bibfnamefont{M.}~\bibnamefont{Voloshin}},
  \bibinfo{journal}{Physical Review D} \textbf{\bibinfo{volume}{73}},
  \bibinfo{pages}{025015} (\bibinfo{year}{2006}).

\bibitem[{\citenamefont{McInnes et~al.}(2018)\citenamefont{McInnes, Healy, and
  Melville}}]{mcinnes2018umap}
\bibinfo{author}{\bibfnamefont{L.}~\bibnamefont{McInnes}},
  \bibinfo{author}{\bibfnamefont{J.}~\bibnamefont{Healy}}, \bibnamefont{and}
  \bibinfo{author}{\bibfnamefont{J.}~\bibnamefont{Melville}},
  \bibinfo{journal}{arXiv preprint arXiv:1802.03426}  (\bibinfo{year}{2018}).

\bibitem[{\citenamefont{Marchenko and
  Pastur}(1967)}]{marchenko1967distribution}
\bibinfo{author}{\bibfnamefont{V.~A.} \bibnamefont{Marchenko}}
  \bibnamefont{and} \bibinfo{author}{\bibfnamefont{L.~A.}
  \bibnamefont{Pastur}}, \bibinfo{journal}{Matematicheskii Sbornik}
  \textbf{\bibinfo{volume}{114}}, \bibinfo{pages}{507} (\bibinfo{year}{1967}).

\end{thebibliography}
\end{document}